\documentclass[12pt]{article}
\usepackage{amssymb,amsmath,epsfig}

\begin{document}
\title{\textbf{Viable and Stable Compact Stars in $f(\mathcal{Q})$ Theory}}
\author{{Muhammad Zeeshan Gul $^1$}\thanks
{mzeeshangul.math@gmail.com}, Shamaila Rani $^{2,~3}$\thanks
{shamailatoor.math@yahoo.com; drshamailarani@cuilahore.edu.pk},\\
{Muhammad Adeel $^3$}\thanks {mr.adimaths@gmail.com} and {Abdul
Jawad $^{2,~3}$}\thanks{jawadab181@yahoo.com; abduljawad@cuilahore.edu.pk}\\
$^1$ Department of Mathematics and Statistics,\\ The University of
Lahore, 54792, Pakistan\\
$^2$ Institute for Theoretical Physics and Cosmology\\
Zhejiang University of Technology, Hangzhou 310023,\\ P. R. China.\\
$^3$ Department of Mathematics, COMSATS University\\ Islamabad,
Lahore-Campus, Lahore-54000, Pakistan.}

\date{}
\maketitle

\begin{abstract}
In this paper, we study the viability and stability of anisotropic
compact stars in the context of $f(\mathcal{Q})$ theory, where
$\mathcal{Q}$ is non-metricity scalar. We use Finch-Skea solutions
to investigate the physical properties of compact stars. To
determine the values of unknown constants, we match internal
spacetime with the exterior region at the boundary surface.
Furthermore, we study the various physical quantities, including
effective matter variables, energy conditions and equation of state
parameters inside the considered compact stars. The equilibrium and
stability states of the proposed compact stars are examined through
the Tolman-Oppenheimer-Volkoff equation, causality condition,
Herrera cracking approach and adiabatic index, respectively. It is
found that viable and stable compact stars exist in $f(\mathcal{Q})$
theory as all the necessary conditions are satisfied.
\end{abstract}


The study of cosmos and its components inspired many researchers in
the last few years due to their mysterious nature. Stars are
considered the basic components of astronomy and the essential
building blocks of galaxies. Fusion processes have significant
effects on the development of stars and planets. The equilibrium
position of stars is maintained through the fusion process if there
is a balance between the inward force of gravity and the outward
pressure. After the consumption of nuclear fuel, the star collapses
and as a result new compact objects like white dwarfs, neutron stars
and black holes are formed depending on their initial mass. Compact
stars have a different nature than ordinary stars as they have large
masses and small radii. These cosmic objects have attracted the
attention of many researchers due to their significant features.
Baade and Zwicky \cite{1} investigated the geometry of compact
objects and proposed the concept of pulsars like Her X-1. After
discovering pulsars, the theory of neutron stars acquired
observational validation \cite{2}. Dev and Gleiser \cite{3} analyzed
the physical behavior of pulsars with different considerations. Mak
and Harko \cite{4} used the mass-radius relationship to analyze the
stability of pulsars. Kalam et al \cite{5} examined the viability
and stability of compact stars using the Karori-Barua technique. The
dynamics of compact objects near the boundary with massless and
massive scalar field are explored in \cite{5a}.

The anisotropy modifies some significant characteristics of
relativistic objects. According to Ruderman \cite{6}, nuclear matter
demonstrates anisotropy if the matter density of relativistic
particles is equivalent to $10^{15}gcm^{-3}$. Due to phase
transition and viscosity, the distribution of matter exhibits
pressure anisotropy \cite{7}. Bowers and Liang \cite{8} examined the
anisotropy of a relativistic sphere and the physical characteristics
of anisotropic pressure. The effects of local anisotropy for
self-gravitating systems have been examined in \cite{9}. The
equilibrium composition and static spherical anisotropic solution
have been studied in \cite{10}. Karori-Barua solutions were used to
analyze the behavior of anisotropic quark stars in \cite{11}. Dourah
and Ray \cite{12} studied the metric solutions for compact stars.
Later, Finch and Skea \cite{13} modified the metric solutions in
four dimensions for anisotropic star models. The Finch-Skea
solutions were used to develop relativistic star models \cite{15}.
Bhar \cite{16} determined the physical characteristics of compact
stars by using the equation of state (EoS) parameter and Finch-Skea
solutions. Anisotropic stellar structures using Finch-Skea
potentials have been examined in \cite{17}. These solutions were
also used to evaluate the anisotropic compact configurations
\cite{18}.

In modified theories of gravity, the study of stellar structures is
a major topic for discussion. Numerous studies on star structures
have been analyzed in the last few decades. Accordingly, the
Symmetric teleparallel gravity, which is also known as
$f(\mathcal{Q})$ is an intriguing theory that has gained attention
in recent years \cite{19}-\cite{21}. The study of $f(\mathcal{Q})$
gravity is the most debatable phenomenon of the current time. Lazkoz
et al \cite{22} established a credible set of limitations on $f(Q)$
gravity, where the polynomial expression of gravity is given as a
function of redshift. Moreover, the $f(\mathcal{Q})$ gravity showed
some fascinating results using observational measurements
\cite{23}-\cite{28}. Furthermore, the study of different cosmic
objects with different matter configurations in the framework of
$f(\mathcal{Q})$ gravity has been discussed in \cite{29}-\cite{32}.
Olmo \cite{33} studied the geometry of compact stellar objects using
polytropic EoS in Palatini $f(R)$ gravity. Arapoglu et al \cite{34}
used barotropic EoS to examine the compactness of pulsars in the
same theory. Zubair et al \cite{35} investigated the viable behavior
of rotating neutron stars in the background of $f(R, T)$ theory.
Mustafa and his collaborators \cite{35a} studied compact spherical
structures with different considerations. Maurya et al \cite{36}
analyzed the effect of charge on the stability of spherical objects
through the Karmarkar condition in $f(G, T)$ gravity. Sharif and Gul
studied the Noether symmetry approach \cite{37}, stability of the
Einstein universe \cite{37a} and dynamics of gravitational collapse
\cite{37b} in modified theory. In the framework of off-diagonal
tetrad, the study of anisotropic strange stars  in $f (\tau, T )$
gravity presented in \cite{37a}. Das et al \cite{38} considered
Finch-Skea geometry to study the viable behavior of pulsars in the
context of Einstein Gauss-Bonnet gravity. Dita et al \cite{39}
studied the characteristics of celestial objects using a modified
Van der Waals EoS in the presence of charge in $f(\mathcal{Q})$
gravity. Recently, the study of observational constraints in
modified $f(Q)$ gravity discussed in \cite{39a} and thermal
fluctuations of compact objects as charged and uncharged BHs in
$f(Q)$ gravity are explored in \cite{39b}. Some people have also
studied the characteristics of celestial objects in different
scenarios of modified gravity and obtained interesting results
\cite{jawad}.

In this article, we analyzed the viability and stability of compact
objects by considering Finch Skea solutions in $f(\mathcal{Q})$
theory. The manuscript is arranged as follows. The basics of
$f(\mathcal{Q})$ gravity with anisotropic matter configuration is
presented in Section \textbf{2}. The geometrical explanation of
metric and the evaluation of unknown constant is given in Section
\textbf{3}. Section \textbf{4} analyzes some physical features to
determine the viability of compact stars. Further, we check the
stability analysis in Section \textbf{5}. Our results are summarized
in section \textbf{6}.

\section{$f(\mathcal{Q})$ Theory: Field Equations}

The corresponding integral action with coupling constant as a unity
is expressed as
\begin{equation}\label{1}
S=\frac{1}{2}\int f(\mathcal{Q})\sqrt{-g}d^4x+\int
L_{m}\sqrt{-g}d^4x,
\end{equation}
where $L_{m}$ represents Lagrangian density of matter and $g$ is
determinant of metric tensor. The non-metricity tensor is defined as
\begin{equation}\label{2}
\mathcal{Q}_{k \eta\xi}=\nabla_{k} g_{\eta\xi}=g_{\xi k}-
\Gamma_{\eta\xi}^{l}g_{lk}-\Gamma_{\eta k}^{l}g_{\xi l},
\end{equation}
where $\nabla_{k}$ covariant derivative and  $\Gamma_{\eta\xi}^{l}$
is affine connection, given by
\begin{equation}\label{3}
\Gamma_{\eta\xi}^{l}=K_{\eta\xi}^{l}+ L_{\eta\xi}^{l},
\end{equation}
where $ L_{\eta\xi}^{l}$ and  $K_{\eta\xi}^{l}$ are deformation and
contortion tensors, respectively, defined as
\begin{eqnarray}\label{4}
L_{\eta\xi}^{l}=\frac{1}{2}\mathcal{Q}_{\eta\xi}^{l}-\mathcal{Q}_{(\eta\xi)}^{l},
\quad K_{\eta\xi}^{l}=\frac{1}{2}T_{\eta\xi}^{l}+ T_{(\eta\xi)}^{l}.
\end{eqnarray}
The antisymmetric component of affine connection reduces to torsion
tensor as $ T_{\eta\xi}^{k} = 2\Gamma_{[\eta\xi]}^{l}$. The super
potential can be written as
\begin{equation}\label{5}
P_{\eta\xi}^{k}=\frac{1}{4}[-\mathcal{Q}_{\eta\xi}^{k}+2
\mathcal{Q}_{(\eta\xi)}^{k}+\mathcal{Q}^{k}g_{\eta\xi}-\tilde{\mathcal{Q}^{k}}g_{\eta\xi}
-\delta_{(\eta\mathcal{Q}_{\xi})}^{k}].
\end{equation}
The non-metricity scalar is expressed as
\begin{equation}\label{6}
\mathcal{Q}=-\mathcal{Q}_{\eta\xi k}P^{\eta\xi k}.
\end{equation}
Variation of action (\ref{1}) corresponding to metric tensor yields
field equations of $f(\mathcal{Q})$ gravity as
\begin{equation}\label{7}
\frac{2}{\sqrt{-g}}\nabla_{k}(\sqrt{-g}
f_{\mathcal{Q}}P_{\eta\xi}^{k})+ \frac{1}{2}g_{\eta\xi} f +
f_{\mathcal{Q}}( P_{\eta
kl}\mathcal{Q}^{kl}_{\xi}-2\mathcal{Q}_{kl\xi} P_{\xi}^{kl}) =
-T_{\eta\xi},
\end{equation}
where $f_{\mathcal{Q}}$ depicts the partial derivative with respect
to non-metricity.

Now, we use the static spherically symmetric spacetime to examine
the stellar structures as
\begin{equation}\label{8}
ds^{2}=-e^{\nu(r)}dt^{2}+e^{\lambda(r)}dr^{2}+d\theta^{2}+\sin^{2}\theta
d\phi^{2}.
\end{equation}
We assume anisotropic matter distribution as
\begin{equation}\label{9}
T_{\eta\xi}=(\rho+P_{t})u_{\eta}u_{\xi}+P_{t}g_{\eta\xi}+(P_{r}-P_{t})\xi_{\eta}
\xi_{\xi},
\end{equation}
where $\rho$, $P_{t}$ and $P_{r}$ depict the energy density,
tangential pressure and radial pressure, respectively. By using
Eq.(\ref{9}), the value of non-metricity becomes
\begin{equation}\label{10}
\mathcal{Q}=-\frac{2e^{-\lambda(r)}(1+r \nu^{'}(r))}{r^{2}},
\end{equation}
where prime is the derivative with respect to radial coordinate. The
resulting field equations turn out to be
\begin{eqnarray}\label{11}
\rho&=&\frac{f(\mathcal{Q})}{2}- f_{\mathcal{Q}}[\mathcal{Q} +
\frac{1}{r^{2}}+ \frac{e^{-\lambda}}{r}(\nu^{'}+\lambda^{'})],
\\\label{12}
P_{r}&=&-\frac{f(\mathcal{Q})}{2}+f_{\mathcal{Q}}[\mathcal{Q} +
\frac{1}{r^{2}}],
\\\label{13}
P_{t}&=&-\frac{f(\mathcal{Q})}{2}+
f_{\mathcal{Q}}[\frac{\mathcal{Q}}{2}-e^{-\lambda}{\frac{\nu^{''}}{2}
+(\frac{\nu^{'}}{4}+\frac{1}{2r})\times(\nu^{'}-\lambda^{'})}],
\\\label{14}
0&=&\frac{cot\theta}{2}\mathcal{Q}^{'}f_{\mathcal{Q}\mathcal{Q}}.
\end{eqnarray}
Solving Eq.(\ref{14}), we have
\begin{equation}\label{15}
f_{\mathcal{Q}\mathcal{Q}}=0\Rightarrow f_{\mathcal{Q}} =
k_{1}\Rightarrow f(\mathcal{Q})= k_{1}\mathcal{Q} + k_{2},
\end{equation}
where $k_{1}$ and $k_{2}$ are integration constants. Now using
Eqs.(\ref{10})-(\ref{15}), the corresponding equations of motion
become
\begin{eqnarray}\label{16}
\rho&=&\frac{1}{2r^{2}}[2k_{1}+ 2e^{-\lambda} k_{1}(r
\lambda^{'}-1)-r^{2}k_{2}],
\\\label{17}
P_{r}&=&\frac{1}{r^{2}}[-2k_{1}+2e^{-\lambda} k_{1}(r
\nu^{'}+1)+r^{2}k_{2}],
\\\label{18}
P_{t}&=&\frac{e^{-\lambda}}{4r}[2e^{\lambda}rk_{2} + k_{1}(2+r
\nu')(\nu'-\lambda')+ 2rk_{1}{\nu''}].
\end{eqnarray}
By using field equations (\ref{16})-(\ref{18}), we obtain
\begin{equation}\label{19}
-\frac{\nu^{'}}{2}(\rho+P_{r})-(P_{r})^{'}+\frac{2}{r}(P_{t}-P_{r})=0.
\end{equation}
This is known as Tolman-Oppenheimer-Volkof (TOV) equation in
$f(\mathcal{Q})$ gravity.

\section{Finch-Skea Solutions and Matching Conditions}

Compact stars are fascinating objects that result from the
gravitational collapse of massive stars. Understanding their
structure, composition and behavior is crucial for advancing our
knowledge of fundamental physics and astrophysics. Different
solutions, including Finch-Skea solutions are considered to describe
the properties of compact stars and their various aspects, such as
density profiles, pressure, and temperature distributions.
Finch-Skea solutions provide insights into the behavior of matter at
extreme conditions in compact stars. These solutions predict
distinct gravitational wave signatures that could be compared with
observations to validate or refine the models. Finch-Skea solutions
provide insights into the interior structure of compact stars,
helping astrophysicists to understand phenomena such as the
formation of quark-gluon plasma or other exotic states of matter.
The Finch-Skea solutions are finite and non-singular, ensuring that
the spacetime is smooth and free from singularities.

The modeling of stellar objects using Finch-Skea solutions has
attained a lot of interest in recent years due to their non-singular
behavior. These solutions are defined as \cite{13}
\begin{eqnarray}\label{20}
e^{\nu}=(A+\frac{1}{2}Br \sqrt{r^{2}C})^{2}, \quad e^{\lambda} =
(1+Cr^{2}).
\end{eqnarray}
where arbitrary constants are denoted by $A$, $B$ and $C$,
respectively. The set of constants can be evaluated by smoothly
matching the inner and outer regions. The Schwarzschild spacetime is
used to investigate the outer geometry of the compact stars as
\begin{equation}\label{21}
ds^{2} = (1-\frac{2M}{r})dt^{2}+
(1-\frac{2M}{r})^{-1}dr^{2}-r^{2}d\theta^{2}-r^{2}\sin^{2}\theta
d\phi^{2},
\end{equation}
where  $M$ and $R$ depict the total mass and radius of the sphere,
respectively. The continuity of metric tensors at the boundary
surface gives
\begin{eqnarray}\label{22}
A=\frac{3M-2R}{2\sqrt{R}\sqrt{R-2M}}, \quad B =
\frac{\sqrt{\frac{M}{R-2M}}\sqrt{R-2M}}{\sqrt{2}R^{\frac{3}{2}}}
\quad C = \frac{2M}{R^{2}(R-2M)}.
\end{eqnarray}
Now, using the Finch Skea metric in Eqs.(\ref{16})-(\ref{18}), we
obtain
\begin{eqnarray}\label{23}
\rho&=&\frac{1}{2r^{2}}\bigg[4 - 2C r^{2}-r^{2}\bigg],
\\\label{24}
P_{r}&=&\frac{1}{2r^{2}}\bigg[-2-2(1+Cr^{2})
(\frac{Br\sqrt{r^{2}C}}{A+\frac{1}{2}Br\sqrt{r^{2}C}}+1)+r^{2}\bigg],
\\\nonumber
P_{t}&=& \frac{-(1+Cr^{2})}{4}\bigg[2(1+Cr^{2})+
(\frac{Br\sqrt{r^{2}C}}{A+\frac{1}{2}Br\sqrt{r^{2}C}}+2)
\\\label{25}&\times&(\frac{Br\sqrt{r^{2}C}}{A+\frac{1}{2}Br\sqrt{r^{2}C}}-\frac{2Cr}{1+Cr^{2}})
+\frac{Ar+\frac{1}{2}Br^{2}\sqrt{r^{2}C}-2BCr^{3}}{(\frac{Br\sqrt{r^{2}C}}{A+\frac{1}{2}Br\sqrt{r^{2}C}})^{2}}\bigg].
\end{eqnarray}
We use green, red, blue, orange, purple, brown, yellow, black, pink
and gray colors for Her X-1, EXO 1785-248, Vela X-1, PSR J1614-2230,
LMC X-4, SMC X-4, PSR J1903+327, 4U 1538-52, 4U 1820-30, Cen X-3
compact stars, respectively for all graphs. Table \textbf{1}
provides the values of unknown constants. Figure \textbf{1} shows
that the metric elements are regular and show positively increasing
behavior as required.
\begin{figure} \epsfig{file=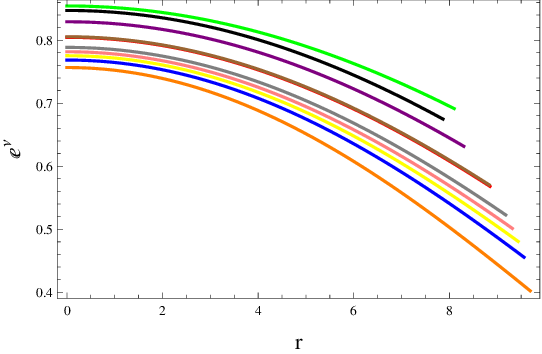,
width=.5\linewidth} \epsfig{file=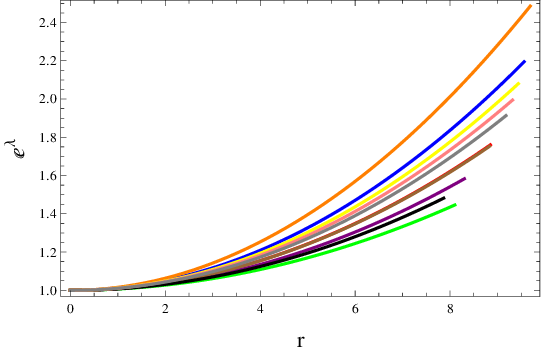, width=.5\linewidth}
\caption{Plot of metric elements versus $\mathrm{r}$.}
\end{figure}
\begin{table}\caption{For the considered
compact star  approximate values of unknown constants.}
\begin{center}
\begin{tabular}{|c|c|c|c|c|c|c|c|}
\hline Star models & $\mathrm{M}(\mathrm{M}_{\odot})$ &
$\mathrm{R}(km)$
& $A$ & $B(km^{-1})$  & $C(km^{-2})$ \\
\hline  Her X-1 & 0.85 & 8.1 & 0.0082 & 0.5725 & 0.1216\\
\hline  EXO 1785-248 & 1.3 & 8.84 &  0.0065 & 0.8586 & 0.1842\\
\hline  Vela X-1 & 1.77 & 9.55 &  0.0076 & 0.9464 & 0.1212\\
\hline  PSR J1614-2230 & 1.97 & 9.70 & 0.0089 & 0.3475 & 0.1854\\
\hline  LMC X-4 & 1.04 & 8.30 & 0.0024 & 0.7867 & 0.2144\\
\hline  SMC X-4 & 1.29 & 8.83 & 0.0051 & 0.2662 & 0.1924\\
\hline  PSR J1903+327 & 1.66 & 9.43 & 0.0026 & 0.5283 & 0.1798\\
\hline  4U 1538-52 & 0.87 & 7.68 & 0.0039 & 0.7249 & 0.2117\\
\hline  4U 1820-30 & 1.58 & 9.31 & 0.0068 & 0.4539 & 0.2097\\
\hline  Cen X-3 & 1.49 & 9.178 & 0.0053 & 0.5850 & 0.2187\\
\hline
\end{tabular}
\end{center}
\end{table}

\section{Physical Attributes}

In this section, we examine the physical characteristics of compact
stars through graphs in the background of $f(\mathcal{Q})$ gravity.
We evaluate the behavior of effective matter variable, anisotropy,
energy condition, mass, compactness, redshift and EoS parameters in
the interior of proposed compact stars.  Further, we use the TOV
equation, sound speed and adiabatic index to analyze the equilibrium
stability state of considered stars.

\subsection{Energy Density and Pressure Components}

Figure \textbf{2} shows that the behavior of energy density, radial
pressure and tangential pressure is positive and decreasing for all
considered compact star candidates. It can also be seen that matter
variables are maximum at the center of stars. Figure \textbf{3}
represents that the radial derivative of energy density, radial and
tangential pressure components are negative, which ensures the
presence of a highly compact picture of the considered compact stars
in the framework of $f(\mathcal{Q})$ gravity.
\begin{figure}
\epsfig{file=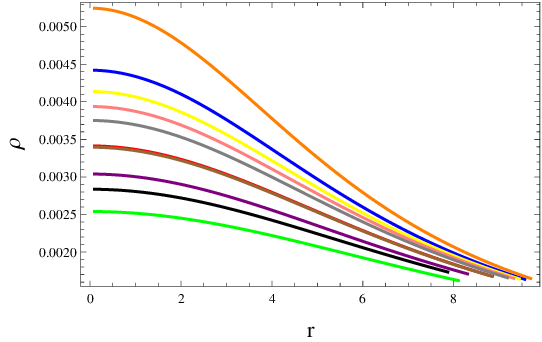,width=.5\linewidth}
\epsfig{file=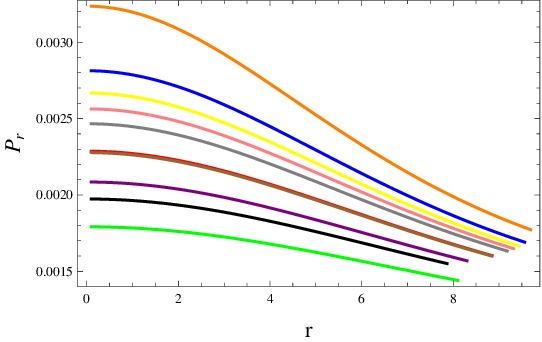,width=.5\linewidth}\center
\epsfig{file=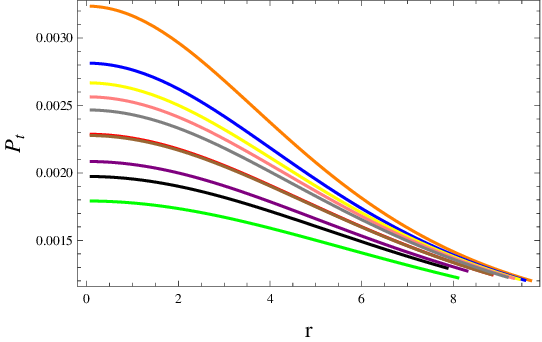,width=.5\linewidth}\caption{Behavior of energy
matter variables versus $\mathrm{r}$.}
\end{figure}
\begin{figure}
\epsfig{file=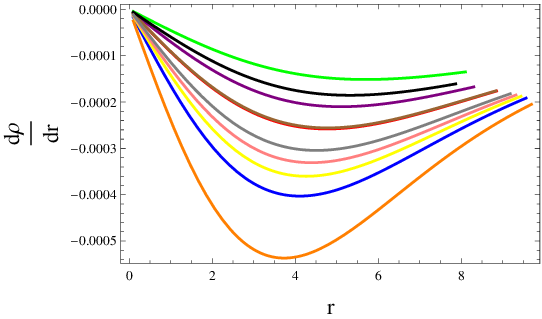,width=.5\linewidth}
\epsfig{file=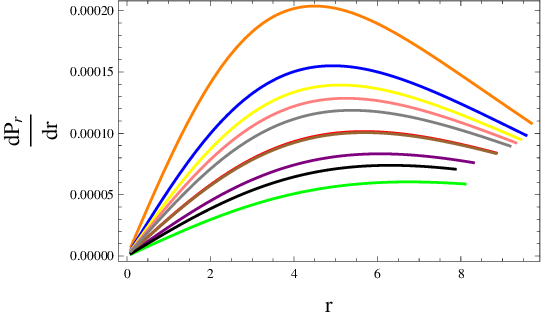,width=.5\linewidth}\center
\epsfig{file=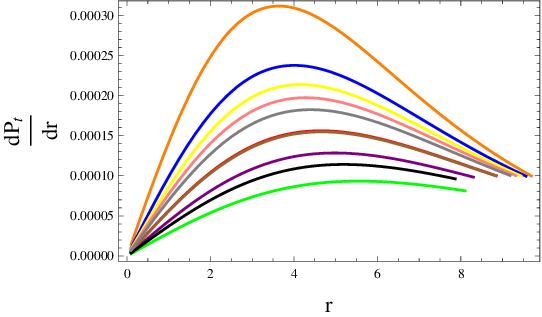,width=.5\linewidth}\caption{Behavior of energy
density and pressure for gradient versus $\mathrm{r}$.}
\end{figure}

\subsection{Anisotropy}

The anisotropy of compact objects can be evaluated  by using
Eqs.(\ref{24}) and (\ref{25}) as
\begin{eqnarray}\nonumber
\Delta &=& P_{t}-P_{r}= \frac{-(1+Cr^{2})}{4}\bigg[2(1+Cr^{2})+
(\frac{Br\sqrt{r^{2}C}}{A+\frac{1}{2}Br\sqrt{r^{2}C}}+2)
\\\nonumber&\times&(\frac{Br\sqrt{r^{2}C}}{A+\frac{1}{2}
Br\sqrt{r^{2}C}}-\frac{2Cr}{1+Cr^{2}})
+\frac{Ar+\frac{1}{2}Br^{2}\sqrt{r^{2}C}-2BCr^{3}}
{(\frac{Br\sqrt{r^{2}C}}{A+\frac{1}{2}Br\sqrt{r^{2}C}})^{2}}\bigg]
\\\nonumber&-&
\frac{1}{2r^{2}}\bigg[-2-2(1+Cr^{2})
(\frac{Br\sqrt{r^{2}C}}{A+\frac{1}{2}Br\sqrt{r^{2}C}}+1)+r^{2}\bigg].
\end{eqnarray}
Anisotropy determines the direction of pressure, i.e., when
$\Delta>0$ the pressure is directed outward and when $\Delta<0$, the
the direction of the pressure is inward. Figure  \textbf{4}
determines that the pressure is in the outward direction as
anisotropy is positive, which is required for compact star
configuration.
\begin{figure}\center
\epsfig{file=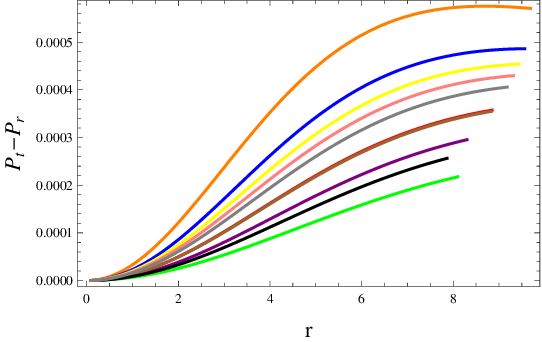,width=.5\linewidth} \caption{Plot of anisotropy
versus $\mathrm{r}$.}
\end{figure}

\subsection{Energy Bounds}

Energy conditions are essential for understanding several
cosmological findings connected to significant gravitational fields.
Due to the vital role of energy bounds, some interesting results
have been published in \cite{40}-\cite{42}. For anisotropic fluid,
$\mathcal{NEC}$ (null energy condition), $\mathcal{WEC}$ (weak
energy condition), $\mathcal{SEC}$ (strong energy condition) and
$\mathcal{DEC}$( dominant energy condition) can be classified as
\begin{itemize}
\item
$\mathcal{NEC}:\rho+P_{r}\geq 0, \quad \rho+P_{t}\geq 0$,
\item
$\mathcal{WEC}:\rho\geq, \quad \rho+P_{r}\geq 0, \quad \rho+P_{t}
\geq 0$,
\item
$\mathcal{DEC}:\rho-P_{r}\geq 0, \quad \rho-P_{t} \geq 0$,
\item
$\mathcal{SEC}:\rho+P_{r}\geq 0, \quad \rho+P_{t}\geq
0,\rho+2P_{t}+P_{r}\geq 0$.
\end{itemize}
Figure \textbf{5} shows that all energy conditions are satisfied for
all star models.
\begin{figure}
\epsfig{file=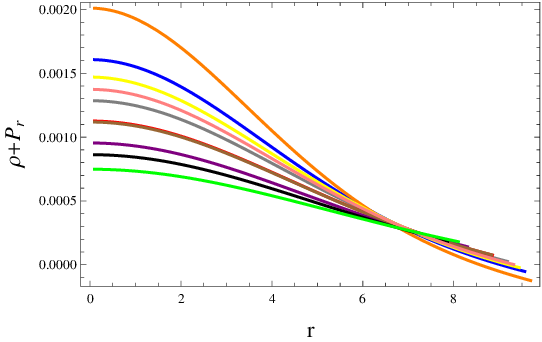,width=.5\linewidth}
\epsfig{file=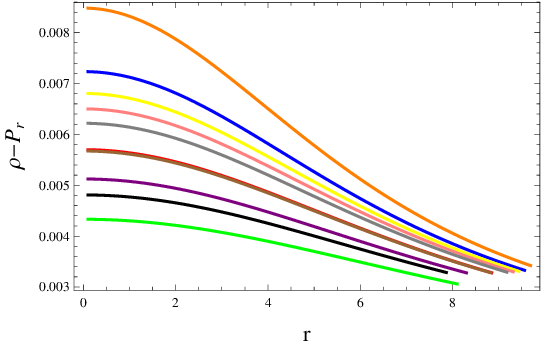,width=.5\linewidth}
\epsfig{file=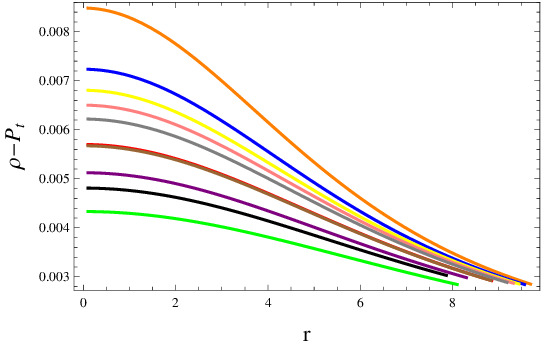,width=.5\linewidth}
\epsfig{file=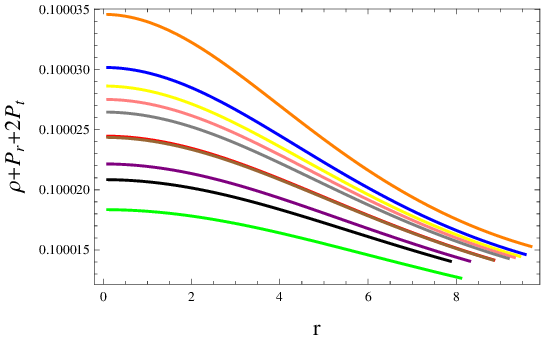,width=.5\linewidth}\caption{Behavior of energy
conditions versus $\mathrm{r}$.}
\end{figure}

\subsection{EoS Parameters}

Here, we investigate the crucial EoS parameters in describing the
relationship between pressure and energy density in various physical
systems. The radial $(\phi_{r}=\frac{P_r}{\rho})$ and transverse
$(\phi_{t}=\frac{P_t}{\rho})$ components must lie in [0,1] for a
physically viable model. The corresponding parameters are expressed
as
\begin{eqnarray}\nonumber
\phi_{r}&=&\bigg[\frac{1}{2r^{2}}\{(\frac{-4B\sqrt{r^{2}C}-
4BCr^{2}\sqrt{r^{2}C}}{A+\frac{1}{2}Br\sqrt{r^{2}C}})
-(\frac{4BCr^{2}\sqrt{r^{2}C}}{A+\frac{1}{2}Br\sqrt{r^{2}C}})+2r\}\bigg]\\\nonumber&\times&
\bigg[\frac{1}{2r^{2}}(4 - 2C r^{2}-r^{2})\bigg]^{-1},
\\\nonumber\phi_{t} &=& \bigg[\{\frac{-(1+Cr^{2})}{4}(2(1+Cr^{2})+
(\frac{Br\sqrt{r^{2}C}}{A+\frac{1}{2}Br\sqrt{r^{2}C}}+2)
\\\nonumber&\times&(\frac{Br\sqrt{r^{2}C}}{A+\frac{1}{2}Br\sqrt{r^{2}C}}
-\frac{2Cr}{1+Cr^{2}}) +\frac{Ar+\frac{1}{2}Br^{2}\sqrt{r^{2}C}-
2BCr^{3}}{(\frac{Br\sqrt{r^{2}C}}{A+\frac{1}{2}Br\sqrt{r^{2}C}})^{2})})\}\bigg]
\\\nonumber&\times&
\bigg[\frac{1}{2r^{2}}(4 - 2C r^{2}-r^{2})\bigg]^{-1}.
\end{eqnarray}
The graphical behavior of $\phi_{r}$ and $\phi_{t}$ for considered
compact star models is given in Figure \textbf{6}, which shows that
EoS parameters satisfy the required condition ($0<\phi_{r}<1$ and
$0<\phi_{t}<1$).
\begin{figure}
\epsfig{file=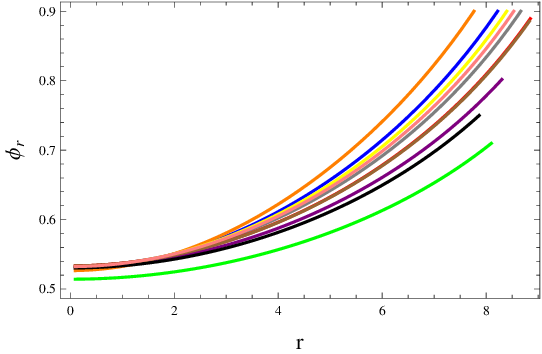,width=.5\linewidth}
\epsfig{file=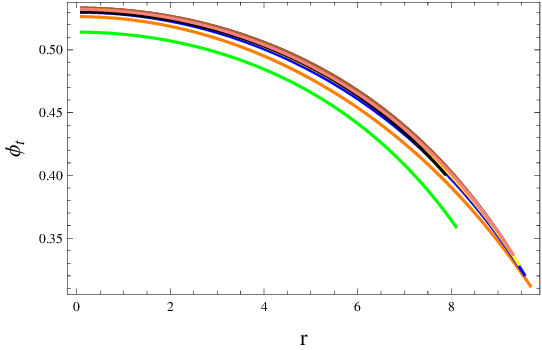,width=.5\linewidth}\caption{ Behavior of EoS
parameters versus $\mathrm{r}$.}
\end{figure}

\subsection{ Mass, Compactness and Redshift}

Mass function for the anisotropic compact stars is given by
\begin{equation}\label{26}
m=\frac{Mr^{3}e^{\frac{M(R^{2}-r^{2})}{R^{2}(2M-R)}}}{R^{2}
(2M-R)+2Mr^{2}e^{\frac{M(R^{2}-r^{2})}{R^{2}(2M-R)}}}.
\end{equation}
Figure \textbf{7} shows that the mass is increasing in a positive
direction and regular at the  center of stars. The compactness
function is essential for examining the viability of compact stars,
expressed as
\begin{equation}\label{27}
\mu(r)=\frac{m(r)}{r}=\frac{Mr^{2}e^{\frac{M(R^{2}-r^{2})}{R^{2}
(2M-R)}}}{R^{2}(2M-R)+2Mr^{2}e^{\frac{M(R^{2}-r^{2})}{R^{2}(2M-R)}}}.
\end{equation}
\begin{figure}\center
\epsfig{file=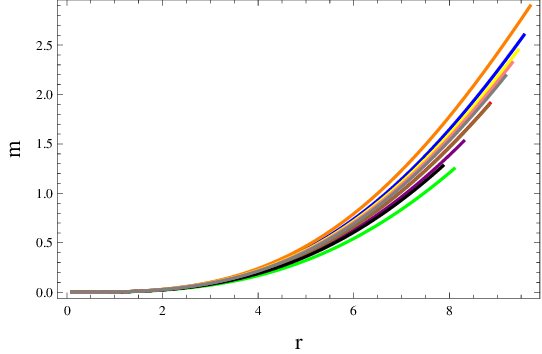,width=.5\linewidth} \caption{Plot of mass
function  versus $\mathrm{r}$.}
\end{figure}
According to Buchdahl \cite{43} compact stellar objects are feasible
if this factor has the limit $\mu(r)<\frac{4}{9}$. The gravitational
redshift  ($ Z = (1-2\mu)^{\frac{-1}{2}}-1$) is considered to be the
key concept to understand the nature of compact objects.
\begin{figure}
\epsfig{file=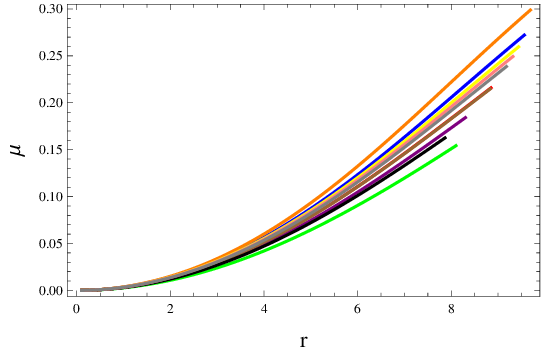,width=.5\linewidth}
\epsfig{file=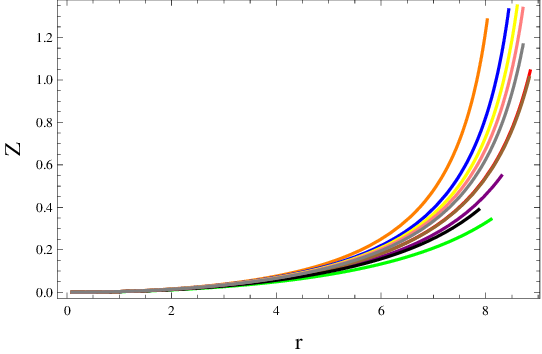,width=.5\linewidth}\caption{ Behavior of
Compactness and redshift functions versus $\mathrm{r}$.}
\end{figure}
Figure \textbf{8} determines that the compactness and redshift lie
in the required limits $(\mu<\frac{4}{9}, Z \leq 5.2)$.

\section{Stability Analysis}

It is more interesting to examine celestial objects that maintain
their stability in the presence of external disturbances. Here, we
use the TOV equation to explore the equilibrium state of the star
candidates and sound speed as well as adiabatic index to check the
stability.

\subsection{Tolman-Oppenheimer-Volkoff Equation}

The TOV equation in the framework of $f(Q)$ is formulated in
Eq.(\ref{19}). This provides information regarding the cosmic
balance as a consequence of the several forces, including the
gravitational force $(F_{g})$, anisotropic force $(F_{a})$ and the
hydrostatic force $(F_{h})$, expressed as
\begin{eqnarray}\nonumber
F_{g}&=& \bigg[-\frac{\nu^{'}}{2}\{ \frac{1}{2r^{2}}(4 - 2C
r^{2}-r^{2})\}\bigg]\\\nonumber&+& \bigg[
\frac{1}{2r^{2}}\{-2-2(1+Cr^{2})
(\frac{Br\sqrt{r^{2}C}}{A+\frac{1}{2}Br\sqrt{r^{2}C}}+1)+r^{2}\}\bigg],
\\\nonumber F_{h}&=&\bigg[\frac{-1}{r^{3}}\{-2-2(1+Cr^{2})
(\frac{Br\sqrt{r^{2}C}}{A+\frac{1}{2}Br\sqrt{r^{2}C}}+1)+r^{2}\}\bigg]
\\\nonumber&+&\bigg[\frac{1}{2r^{2}}\{(\frac{-4B\sqrt{r^{2}C}-
4BCr^{2}\sqrt{r^{2}C}}{A+\frac{1}{2}Br\sqrt{r^{2}C}})
-(\frac{4BCr^{2}\sqrt{r^{2}C}}{A+\frac{1}{2}Br\sqrt{r^{2}C}})+2r\}\bigg],
\\\nonumber F_{a}&=&\frac{2}{r}\bigg[\{\frac{-(1+Cr^{2})}{4}(2(1+Cr^{2})+
(\frac{Br\sqrt{r^{2}C}}{A+\frac{1}{2}Br\sqrt{r^{2}C}}+2)
\\\nonumber&\times&(\frac{Br\sqrt{r^{2}C}}{A+\frac{1}{2}Br\sqrt{r^{2}C}}
-\frac{2Cr}{1+Cr^{2}}) +\frac{Ar+\frac{1}{2}Br^{2}\sqrt{r^{2}C}-
2BCr^{3}}{(\frac{Br\sqrt{r^{2}C}}{A+\frac{1}{2}Br\sqrt{r^{2}C}})^{2})})\}\bigg]
\\\nonumber&-&\bigg[\frac{1}{2r^{2}}\{(\frac{-4B\sqrt{r^{2}C}-
4BCr^{2}\sqrt{r^{2}C}}{A+\frac{1}{2}Br\sqrt{r^{2}C}})
-(\frac{4BCr^{2}\sqrt{r^{2}C}}{A+\frac{1}{2}Br\sqrt{r^{2}C}})+2r\}\bigg].
\end{eqnarray}
Figure \textbf{9} shows that the our system is in the equilibrium
state as the sum of all forces is zero $(F_{g}+F_{h}+F_{a}=0)$.
\begin{figure}
\epsfig{file=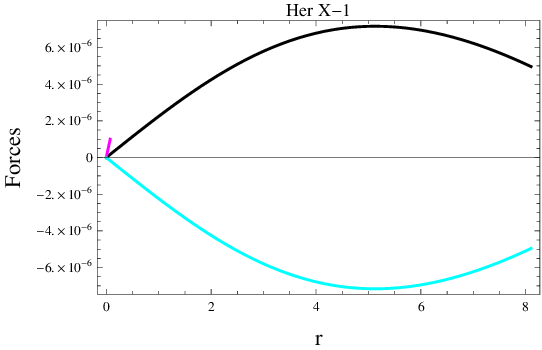,width=.5\linewidth}
\epsfig{file=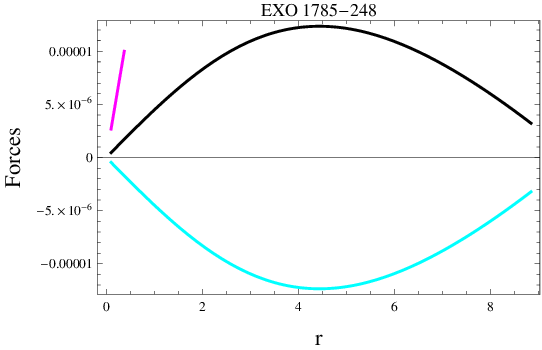,width=.5\linewidth}
\epsfig{file=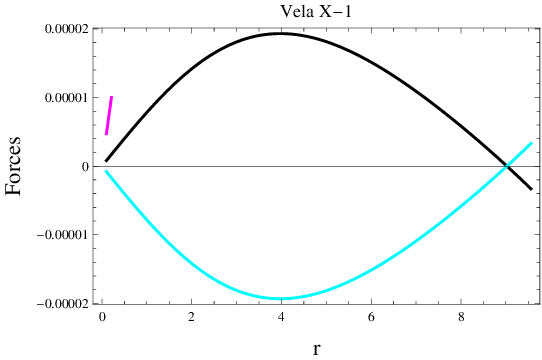,width=.5\linewidth}
\epsfig{file=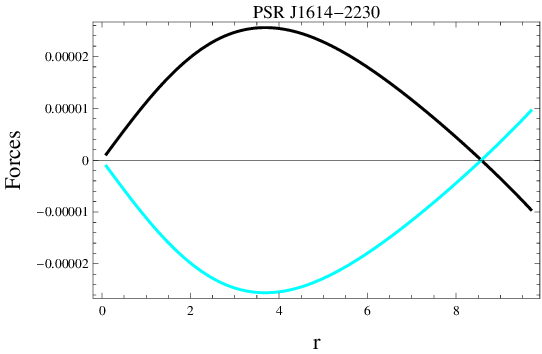,width=.5\linewidth}
\epsfig{file=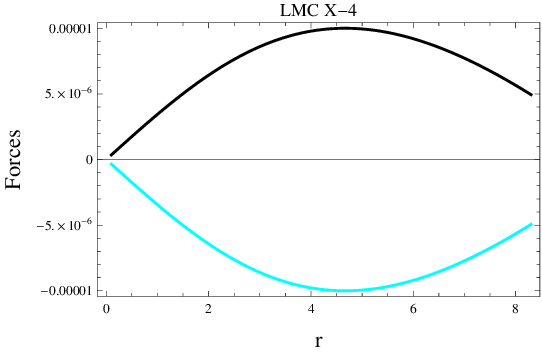,width=.5\linewidth}
\epsfig{file=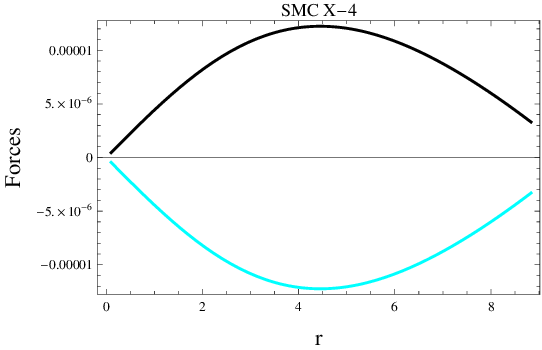,width=.5\linewidth}
\epsfig{file=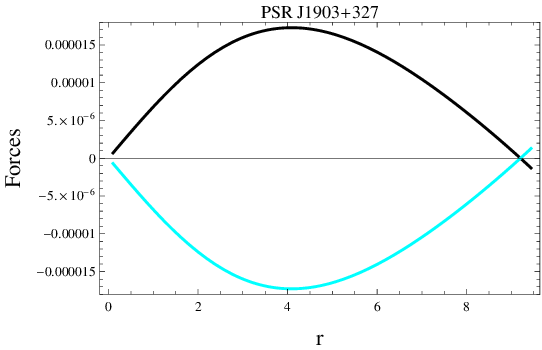,width=.5\linewidth}
\epsfig{file=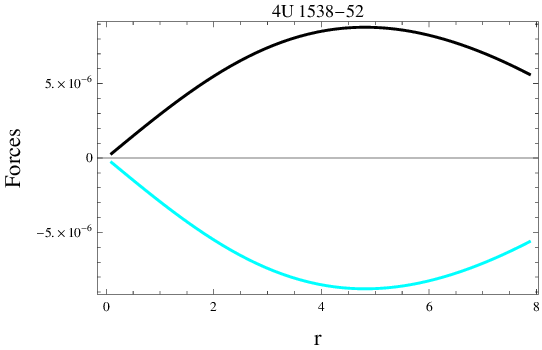,width=.5\linewidth}
\epsfig{file=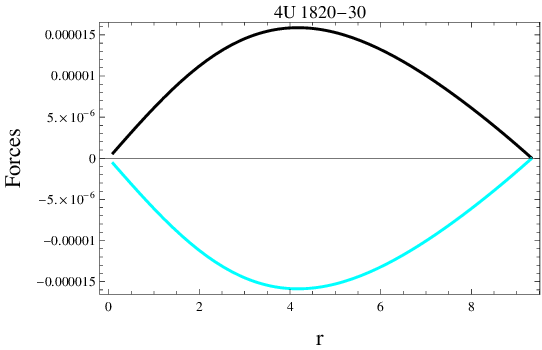,width=.5\linewidth}
\epsfig{file=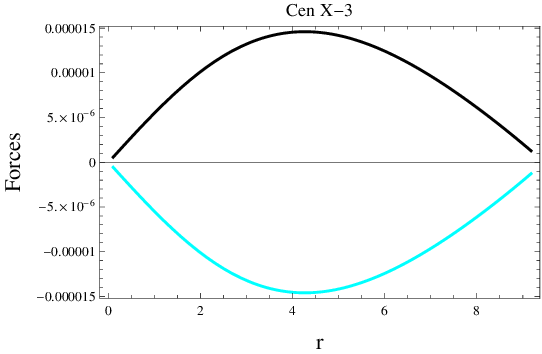,width=.5\linewidth}\caption{Behavior of TOV
equation versus $\mathrm{r}$.}
\end{figure}

\subsection{Sound Speed}

The development of cracking technique has been explored spherically
for compact objects using various methodologies \cite{44}-\cite{46}.
The radial and transverse sound speed components, represented as
$\nu^{2}_{st}$ and $\nu^{2}_{st}$ are used to determine the
stability of compact star candidates. The expressions for speed of
sounds are given as follows
\begin{eqnarray}\nonumber
\nu^{2}_{sr}&=& \bigg[\frac{-1}{r^{3}}\{-2-2(1+Cr^{2})
(\frac{Br\sqrt{r^{2}C}}{A+\frac{1}{2}Br\sqrt{r^{2}C}}+1)+r^{2}\}\bigg]
\\\nonumber&+&\bigg[\frac{1}{2r^{2}}\{(\frac{-4B\sqrt{r^{2}C}-
4BCr^{2}\sqrt{r^{2}C}}{A+\frac{1}{2}Br\sqrt{r^{2}C}})
-(\frac{4BCr^{2}\sqrt{r^{2}C}}{A+\frac{1}{2}Br\sqrt{r^{2}C}})+2r\}\bigg]
\\\nonumber&\times&\bigg[\frac{1}{2r^{2}}(-4Cr-2r)-\frac{1}{r^{3}}(4-2Cr^{2}-r^{2})\bigg]^{-1},
\\\nonumber\nu^{2}_{st}&=& \bigg[\{\frac{-(1+Cr^{2})}{4}(2(1+Cr^{2})+
(\frac{Br\sqrt{r^{2}C}}{A+\frac{1}{2}Br\sqrt{r^{2}C}}+2)
\\\nonumber&\times&(\frac{Br\sqrt{r^{2}C}}{A+\frac{1}{2}Br\sqrt{r^{2}C}}
-\frac{2Cr}{1+Cr^{2}}) +\frac{Ar+\frac{1}{2}Br^{2}\sqrt{r^{2}C}-
2BCr^{3}}{(\frac{Br\sqrt{r^{2}C}}{A+\frac{1}{2}Br\sqrt{r^{2}C}})^{2})})\}\bigg]
\\\nonumber&\times&\bigg[\frac{1}{2r^{2}}(-4Cr-2r)-\frac{1}{r^{3}}(4-2Cr^{2}-r^{2})\bigg]^{-1}.
\end{eqnarray}
The causality condition in the case of anisotropic matter
configuration is defined as
\begin{eqnarray}\nonumber
0\leq v^{2}_{sr}\leq 1, \quad 0\leq v_{st}^{2}\leq 1.
\end{eqnarray}
The range of radial sound speed and transverse sound speed must
satisfy the following inequality $0\leq |v^{2}_{sr}-v^{2}_{st}|\leq
1$ for stable compact stars. We observed that the desired conditions
are fulfilled as shown in Figures \textbf{10} and \textbf{11}. This
proves the stability of our compact star models in the framework of
$f(\mathcal{Q})$ gravity.
\begin{figure}
\epsfig{file=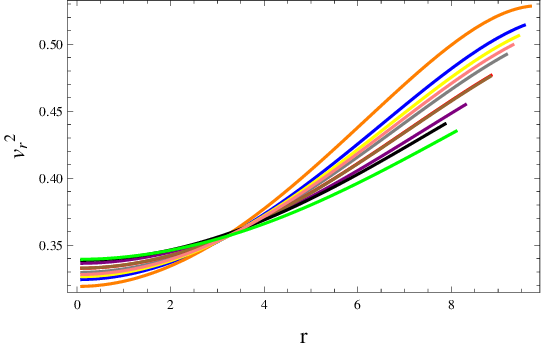,width=.5\linewidth}
\epsfig{file=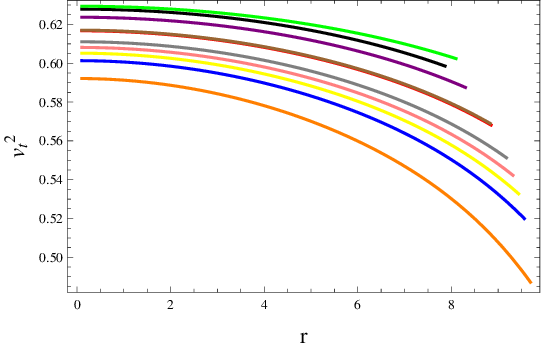,width=.5\linewidth}\caption{Behavior of
components of sound speed versus $\mathrm{r}$.}
\end{figure}
\begin{figure}\center
\epsfig{file=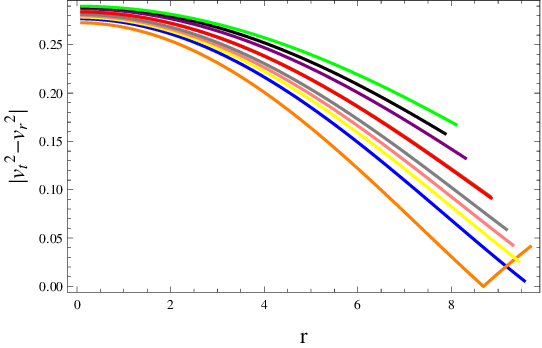,width=.5\linewidth}\caption{Behavior of $|
\nu^{2}_{t}-\nu^{2}_{r}|$
 versus $\mathrm{r}$.}
\end{figure}

\subsection{Adiabatic Index}

The another alternative technique to investigate the stability of
compact stars is adiabatic index. The radial and transverse
components of adiabatic index are defined as
\begin{eqnarray}\label{28}
\Gamma_{r}= \frac{P_{r}+ \rho}{P_{r}} \frac{dP_{r}}{d\rho}=
\frac{P_{r}+ \rho}{P_{r}}\nu^{2}_{sr}, \quad \Gamma_{t} =
\frac{P_{t}+ \rho}{P_{t}} \frac{dP_{t}}{d\rho}= \frac{P_{t}+
\rho}{P_{t}}\nu^{2}_{st}.
\end{eqnarray}
Using Eqs.(\ref{23})-(\ref{25}), we have
\begin{eqnarray}\nonumber
\Gamma_{r}&=& \bigg[\{\frac{1}{2r^{2}}(4 - 2C r^{2}-r^{2})\}\bigg] +
\bigg[ \frac{1}{2r^{2}}\{-2-2(1+Cr^{2})
(\frac{Br\sqrt{r^{2}C}}{A+\frac{1}{2}Br\sqrt{r^{2}C}}+1)\\\nonumber&+&r^{2}\}\bigg]
\times\bigg[ \frac{1}{2r^{2}}\{-2-2(1+Cr^{2})
(\frac{Br\sqrt{r^{2}C}}{A+\frac{1}{2}Br\sqrt{r^{2}C}}+1)+r^{2}\}\bigg]^{-1}
\\\nonumber&\times&\bigg[\frac{-1}{r^{3}}\{-2-2(1+Cr^{2})
(\frac{Br\sqrt{r^{2}C}}{A+\frac{1}{2}Br\sqrt{r^{2}C}}+1)+r^{2}\}\bigg]
\\\nonumber&+&\bigg[\frac{1}{2r^{2}}\{(\frac{-4B\sqrt{r^{2}C}-
4BCr^{2}\sqrt{r^{2}C}}{A+\frac{1}{2}Br\sqrt{r^{2}C}})
-(\frac{4BCr^{2}\sqrt{r^{2}C}}{A+\frac{1}{2}Br\sqrt{r^{2}C}})+2r\}\bigg]
\\\nonumber&\times&\bigg[\frac{1}{2r^{2}}(-4Cr-2r)-\frac{1}{r^{3}}(4-2Cr^{2}-r^{2})\bigg]^{-1},
\\\nonumber\Gamma_{t}&=&\bigg[\{\frac{1}{2r^{2}}(4 - 2C r^{2}-r^{2})\}\bigg]
+ \bigg[\{\frac{-(1+Cr^{2})}{4}(2(1+Cr^{2})+
(\frac{Br\sqrt{r^{2}C}}{A+\frac{1}{2}Br\sqrt{r^{2}C}}+2)
\\\nonumber&\times&(\frac{Br\sqrt{r^{2}C}}{A+\frac{1}{2}Br\sqrt{r^{2}C}}
-\frac{2Cr}{1+Cr^{2}}) +\frac{Ar+\frac{1}{2}Br^{2}\sqrt{r^{2}C}-
2BCr^{3}}{(\frac{Br\sqrt{r^{2}C}}{A+\frac{1}{2}Br\sqrt{r^{2}C}})^{2})})\}\bigg]
\\\nonumber&\times&\bigg[\{\frac{-(1+Cr^{2})}{4}(2(1+Cr^{2})+
(\frac{Br\sqrt{r^{2}C}}{A+\frac{1}{2}Br\sqrt{r^{2}C}}+2)
\\\nonumber&\times&(\frac{Br\sqrt{r^{2}C}}{A+\frac{1}{2}Br\sqrt{r^{2}C}}
-\frac{2Cr}{1+Cr^{2}}) +\frac{Ar+\frac{1}{2}Br^{2}\sqrt{r^{2}C}-
2BCr^{3}}{(\frac{Br\sqrt{r^{2}C}}{A+\frac{1}{2}Br\sqrt{r^{2}C}})^{2})})\}\bigg]^{-1}
\\\nonumber&\times& \bigg[\{\frac{-(1+Cr^{2})}{4}(2(1+Cr^{2})+
(\frac{Br\sqrt{r^{2}C}}{A+\frac{1}{2}Br\sqrt{r^{2}C}}+2)
\\\nonumber&\times&(\frac{Br\sqrt{r^{2}C}}{A+\frac{1}{2}Br\sqrt{r^{2}C}}
-\frac{2Cr}{1+Cr^{2}}) +\frac{Ar+\frac{1}{2}Br^{2}\sqrt{r^{2}C}-
2BCr^{3}}{(\frac{Br\sqrt{r^{2}C}}{A+\frac{1}{2}Br\sqrt{r^{2}C}})^{2})})\}\bigg]
\\\nonumber&\times&\bigg[\frac{1}{2r^{2}}(-4Cr-2r)-\frac{1}{r^{3}}(4-2Cr^{2}-r^{2})\bigg]^{-1}.
\end{eqnarray}
According to adiabatic index criteria, the system is stable if
$\Gamma>\frac{4}{3}$ otherwise, it is unstable \cite{47}. Figure
\textbf{12} shows that our system is stable in the presence of
modified terms corresponding to all considered compact stars.
\begin{figure}
\epsfig{file=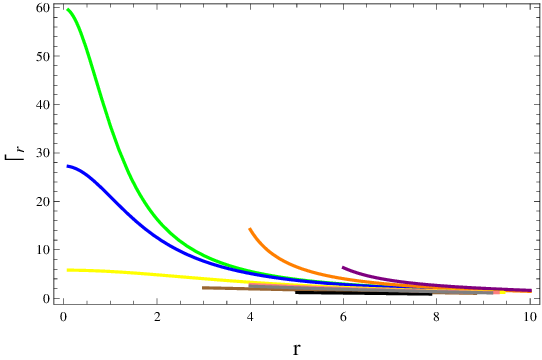,width=.5\linewidth}
\epsfig{file=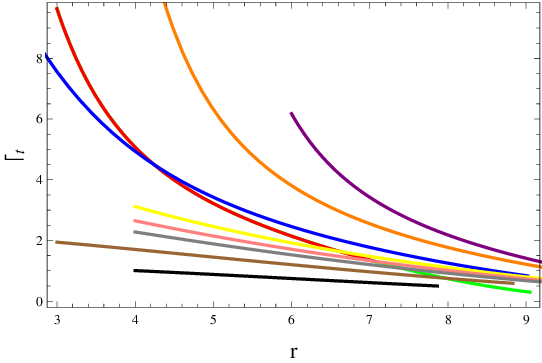,width=.5\linewidth}\caption{Behavior of
adiabatic index versus $\mathrm{r}$.}
\end{figure}

\section{Final Remarks}

In this paper, we have examined the viability and stability of
compact stars in the background of $f(\mathcal{Q})$ theory. To
evaluate the graphical characteristics of compact stars, we
formulate the functional form as
$f(\mathcal{Q})=k_{1}\mathcal{Q}+k_{2}$. The main results are given
by
\begin{itemize}
\item
The graphical behavior of energy density versus radial coordinate
depicts that energy density approaches to the maximum value when
$r\rightarrow 0$ as shown in Figure \textbf{2}. We have noted the
same behavior for $P_{t}$ and $P_{r}$ that is positive and
decreasing. The radial derivative of the energy density and pressure
components are negative for considered compact star candidates as
shown in Figure \textbf{3}.
\item
The anisotropy for the compact star candidates is directed outward
as shown in Figure \textbf{4}.
\item
Energy conditions for compact star candidates are positive, which
ensure the presence of normal matter in the proposed compact stars
as shown in Figure \textbf{5}.
\item
It is examined that the EoS parameters satisfy the required bound
for radial component $ 0< \phi_{r} <1$ and tangential component $ 0<
\phi_{t} <1$ as given in Figure \textbf{6}.
\item
There is a direct relation between the mass and radius of the
compact stars, suggesting that the mass function remains regular at
the center of the compact stars (Figure \textbf{7}).
\item
Compactness and redshift functions satisfied the required limits in
the presence of modified terms as presented in Figure \textbf{8}.
\item
It is found that the proposed compact stars are stable as all the
necessary conditions are fulfilled as shown in Figures
\textbf{9}-\textbf{12}.
\end{itemize}
It is noteworthy to mention here that we have obtained more stable
anisotropic stellar structures due to the presence of
$f(\mathcal{Q})$ terms as compared to GR and other modified theories
\cite{48}-\cite{50}. We can conclude that viable and stable compact
stars exist in this modified theory.

\section*{Acknowledgement}

Shamaila Rani and Abdul Jawad are thankful to ITPC and Zheijiang
University of Technology for providing the postdoctoral opportunity.

\end{document}